\documentclass[fleqn,usenatbib]{mnras}

\usepackage{newtxtext,newtxmath}

\usepackage[T1]{fontenc}

\DeclareRobustCommand{\VAN}[3]{#2}
\let\VANthebibliography\thebibliography
\def\thebibliography{\DeclareRobustCommand{\VAN}[3]{##3}\VANthebibliography}

\usepackage{graphicx}	

\usepackage{amsmath}	
\usepackage{amssymb}	

\title[exposure time bias]{Photometric light curve studies: potential bias induced by exposure time}

\author[Liu D. \& Fan Z.]{
Dezi Liu $^{1}$\thanks{E-mail: adzliu@ynu.edu.cn (SWIFAR)}, 
Zuhui Fan $^{1}$\thanks{E-mail: zuhuifan@ynu.edu.cn (SWIFAR)}
\\
$^{1}$South-Western Institute For Astronomy Research, Yunnan University,  Kunming 650500, China\\
}

\date{Accepted XXX. Received YYY; in original form ZZZ}

\pubyear{2025}

\begin{document}
\label{firstpage}
\pagerange{\pageref{firstpage}--\pageref{lastpage}}
\maketitle

\begin{abstract}
In photometric observations, the flux averaged over the preset exposure time  is usually used as the representation of an object's true flux at the middle of the exposure interval. For the study of transients and variables, it is also the default manner to build the light curves. In this work, we investigate the effect of this common practice on quantifying the photometric light curves. Our analysis shows that the flux averaged over the exposure time is not  necessarily identical to the true flux so that potential bias may be introduced. The overall profile of the true light curve tends to be flattened by the exposure time. In addition, it is found that  the peak position and photometric color can also be altered. We then discuss the impacts of the bias induced by exposure time on the light curves of stellar flares, periodic stars, and active galactic nuclei (AGNs). The bias can lead to an underestimate of the total fluxes of stellar flares which has been noticed in the observational data. For periodic stars that follow a sinusoidal light curve, the bias does not affect the period and peak position, but can result in the peak flux being underestimated. Meanwhile, the bias can result in steeper structure function at short timescales for AGN light curves. To obtain unbiased physical parameter estimates from the light curves, our analysis indicates that it is essential to account for this bias, particularly for transients and variables with very short timescales.
\end{abstract}




\section{Introduction}
Many astronomical objects in the Universe exhibit a variety of dynamic phenomena, often manifesting as variations in brightness or dramatic explosions over different timescales. Based on the types of variability, these objects can be  broadly categorized into two groups: transients and variables \citep{2009arXiv0912.0201L, 2023PASP..135j5002H}. Transients are objects whose brightness undergoes a sudden change and lasts for only a limited span of time, ranging from milliseconds to years. Examples include supernovae, tidal disruption events, and stellar flares \citep[e.g.,][]{2000ARA&A..38..191H, 2009ARA&A..47...63S, 2011MNRAS.410..359L, 2024LRSP...21....1K}. In contrast,  variables are generally referred to as objects that show persistent brightness variation over various timescales without significantly  changing their physical nature, such as pulsating stars and eclipsing binary stars \citep{2008JPhCS.118a2010E}. A large number of observations have indicated that these dynamic phenomena are prevalent in the Universe. The datasets obtained  from these observations are constantly updating our understanding of the physical mechanisms concerning the structure, composition and internal activity of these objects \citep{2021ApJ...908....4V, 2023A&A...674A..13E,2024arXiv240904346R}.

The flux variation of these objects under sequential observations is generally referred to as light curve, which is a time series  of discrete measurements. To extract the statistical metrics (e.g., amplitudes, colors, etc.) that describe the morphological and temporal properties of the light curve without bias, accurate measurements are very important. In regard to the observations, however, many factors can affect the measurement accuracy, e.g. the observing strategies, facility status, weather conditions, data processing methods, and so forth. The temporal resolution, as a potential bias source, is tightly related to the observing strategies. Generally, the temporal resolution consists of two aspects: observing cadence (or sampling rate) and exposure time. The observing cadence defines the time interval between two sequential exposures, while the exposure time is the duration that a detector is uninterruptedly exposed to the target object or a patch of sky. The cadence and exposure time are observationally coupled. High cadence observation usually corresponds to short  single exposure time, and vice versa. In most observations, the cadence includes not only irregular diurnal, lunar, and seasonal gaps but also additional overheads required for routine activities such as filter changes, detector readout, and telescope slew \citep{2018PASP..130f4505T, LSST2019}. Consequently, the cadence is typically uneven and longer than the exposure time. Notable exceptions are observations from missions like \textit{Kepler} \citep{2010Sci...327..977B} and \textit{TESS} \citep{2015JATIS...1a4003R}, which exhibit minimal overheads due to their specialized detector designs. In these cases, the exposure time can be considered virtually identical to the cadence\footnote{In this paper, we use the term exposure time instead of the commonly used cadence for the  \textit{Kepler} and \textit{TESS} light curve.}.

The cadence and exposure time can impose different effects on the observed light curve. The cadence may induce shape deviation from the true light curve due to the discreteness of time sampling. As mentioned above, the cadence is usually unevenly spaced in observation, and can lead to remarkable aliasing in analyzing the periodic light curve \citep{2018ApJS..236...16V}. As for the exposure time, the photometric flux measured within a specified exposure time is an integral value. This  makes the variation within the exposure smeared out so that the observed shape of the light curve may also deviate from the true case. The deviation is expected to depend on the profile of the true light curve and the duration of an exposure time.  Recently, several studies compared the light curves of stellar flares observed with both long exposure time and short exposure time, and concluded that systematic differences exist on the derived parameters \citep{2018ApJ...859...87Y,2022ApJ...926..204H}. It was found that the flare energy and peak amplitude were underestimated by 25\% and 60\%, respectively, while the flare duration overestimated by 50\% for the long exposure light curves. 

With the development of time-domain astronomy, accurate light curve measurements, which are the basis of further study the underlying astrophysics, become crucial. However, few works have discussed the potential  bias induced by the temporal resolution in terms of general light curve analysis. Ideally, as discussed above, observations with short exposure and high cadence are desirable to extract the accurate light curve without introducing significant bias. For bright objects, that is sometimes feasible. But for faint objects, it is not always achievable if high signal-to-noise ratio is required. In this case, exposure times of several minutes or longer are usually applied for these objects. Therefore, understanding any potential systematic effects of temporal resolution on quantifying the light curve is important, particularly for upcoming large sky surveys, e.g. the Vera Rubin Observatory Legacy Survey of Space and Time \citep[LSST;][]{LSST2019}, which are expected to observe more transients and variables with  unprecedented accuracy \citep{2023PASP..135j5002H}.

Since the cadence and exposure time are closely related, in this work, we mainly focus on the study of the impacts of exposure time on quantifying the light curve. The paper is structured as follows: in Section~\ref{sec:biasMath}, we present the bias induced by the exposure time with detailed mathematical analysis. Three examples are then discussed in Section~\ref{sec:cases}. Finally, in Section~\ref{sec:conc}, brief discussions and conclusions are given. In this work, we do not consider noise sources in observational data, such as shoot noise or other stochastic sources resulting from the instruments and measurements.

\section{Bias induced by exposure time}\label{sec:biasMath}

\begin{figure}
    \centering
    \includegraphics[width=1\linewidth]{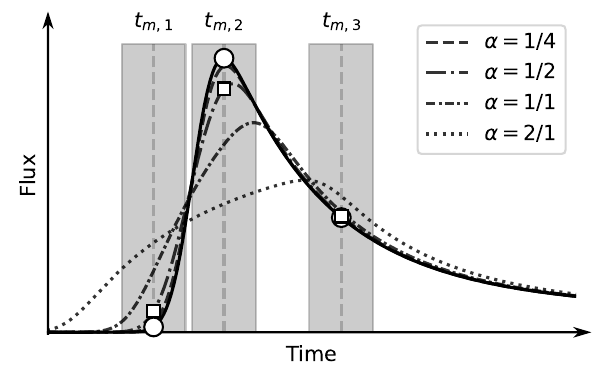}
    \includegraphics[width=1\linewidth]{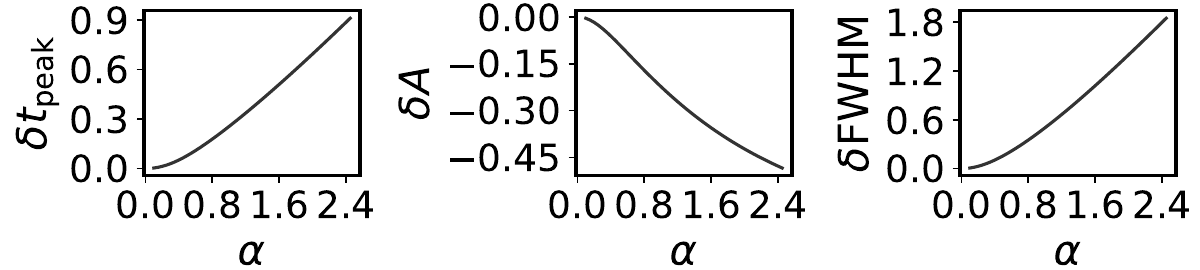}
    \caption{Top panel: a sketch map to show the $\Delta{t}$-bias. The solid line is an empirical light curve  template of  white-light stellar flares. Other  lines represent the time-averaged light curves. Three vertical dashed lines illustrate an example pattern of cadence at times $t_{m,1}$, $t_{m,2}$, and $t_{m,3}$. The width of each shaded region is equal to the exposure time $\Delta{t}=\mathrm{FWHM}/2$. The circles represent the resulting discrete light curve for our flare template, while the squares represent the discrete light curve with $\alpha=1/2$ after incorporating $\Delta{t}$-bias. Bottom panel:  the differences of the profile parameters ($t_\mathrm{peak}$,  $A$, and FWHM) relative to those of the template as a function of exposure time.}
    \label{fig:tbias_demo}
\end{figure}

In photometric observations, the flux of an astronomical object needs to be measured in a specified time interval $[t_{1}, \,t_{2}]$, usually referred to as exposure time, under a given filter. In most light curve analyses, the flux averaged over the exposure time is generally regarded as the flux observed at the middle of the interval $t_{m}=(t_{1}+{t_{2}})/2$. However, this intuitive assumption has never been carefully investigated. In this section, we will explore this assumption with detailed mathematical calculations.

To start with, we assume that the spectral energy distribution $s_{\nu}(t)$ of a given astronomical object, in unit of erg\,s$^{-1}$\,cm$^{-2}$\,Hz$^{-1}$, varies as a function of time $t$. In the meantime, $s_{\nu}(t)$ itself is a function of frequency $\nu$. The true flux of the object observed at $t_{m}$ is the integral of $s_{\nu}(t_{m})$ and the filter transmission curve $R_{\nu}$, which is 
\begin{equation}\label{eq:cflux}
    f(t_{m}) = Z_{f} \int_{0}^{+\infty}s_{\nu}(t_{m})R_{\nu}\frac{\mathrm{d}\nu}{\nu},
\end{equation}
where $Z_{f}$ is the flux zeropoint determined by the telescope and detector characteristics \citep{2018AJ....155...41B}. We note that the filter transmission curve $R_{\nu}$  incorporates contributions from multiple causes, including  the Earth's atmosphere, telescope optics, filter glass, and detector \citep[for a detailed definition, see, e.g.,][]{2022arXiv220600989H}.
The magnitude is then defined as 
\begin{equation}
    m({t_{m}}) = -2.5\log_{10}\left[{\frac{\int_{0}^{+\infty}s_{\nu}(t_{m})R_{\nu}\frac{\mathrm{d}\nu}{\nu}}{\int_{0}^{+\infty}g_{\nu}^\mathrm{ref}R_{\nu}\frac{\mathrm{d}\nu}{\nu}}}\right], \nonumber
\end{equation}
where $g_{\nu}^\mathrm{ref}$ is the time-independent spectral energy  distribution of a standard reference source, which is the spectrum of Vega for Vega magnitude, or a hypothetical constant source with $g_{\nu}^\mathrm{ref}=3631$\,Jy 
at all frequencies for AB magnitude \citep{1983ApJ...266..713O}. Similarly, the time-averaged flux in the time interval [$t_{1}$, $t_{2}$] can be calculated through  
\begin{align}\label{eq:dtFlux}
    f_{s}(t_{m}) &= Z_{f}\frac{1}{\Delta{t}}\int_{t_{m}-\Delta{t}/2}^{t_{m}+\Delta{t}/2}\mathrm{d}t\,\int_{0}^{+\infty}s_{\nu}(t)R_{\nu}\frac{\mathrm{d}\nu}{\nu} \nonumber \\
    \nonumber \\
    &= \frac{1}{\Delta{t}}\int_{t_{m}-\Delta{t}/2}^{t_{m}+\Delta{t}/2} f(t) \mathrm{d}t,
\end{align}
where the time duration $\Delta{t}$, defined as $\Delta{t} =  t_{2} - t_{1}$, is the exposure time.

In observation, we can never measure the true flux $f(t_{m})$ because any facility needs a given time duration to complete an exposure, i.e. $\Delta{t}>0$. Instead, the time-averaged flux  $f_{s}(t_{m})$ is the direct observable. When assembling the light curve of an astronomical object and performing quantitative analysis, we intuitively use $f_{s}(t_{m})$ as a representation of the potentially real $f(t_{m})$. Apparently, when the object is stable and its flux does not vary with time, the true flux and time-averaged flux are equal, i.e $f(t_{m}) = f_{s}(t_{m})$.  Generally, however,  for variables or transients $f_{s}(t_{m})$ and $f(t_{m})$  are not necessarily identical. As a result, this assumption may lead to systematic bias in interpreting the light curves. In the following analysis, we denote this potential bias as exposure time bias ($\Delta{t}$-bias). 

Firstly, under the limit of $\Delta{t}$ approaching to zero the fundamental calculus theorem tells us that 
\begin{equation}\label{eq:biasLim}
f(t_{m}) = \lim_{\Delta{t} \rightarrow 0} f_{s}(t_{m}).
\end{equation}
Or equivalently, if we define $\omega_{t}$ as the typical timescale of $f(t_{m})$, e.g. the full width at half-maximum (FWHM), and a scale parameter $\alpha \equiv  \Delta{t}/\omega_{t}$, equation~(\ref{eq:biasLim}) is also satisfied under the limit of $\alpha$ approaching to zero.
Therefore, it is expected that $f_{s}(t_{m})$ and $f(t_{m})$ will be very close if the exposure time ($\Delta{t}$) or the relative exposure time ($\alpha$), when compared to the timescale of the light curve, is short.

To quantify the $\Delta{t}$-bias in more general case, the simple way is to compare the difference between the two fluxes, which can be defined as 
\begin{equation}\label{eq:diff}
\Delta{f(t_{m})} = f_{s}(t_{m}) - f(t_{m}) = \frac{1}{\Delta{t}}\int_{t_{m}-\Delta{t}/2}^{t_{m}+\Delta{t}/2}[f(t) - f(t_{m})]\mathrm{d}t.
\end{equation} 
For a given light curve, the flux variations typically display sequential brightening and darkening phases, such as the periodic stars and various transients. Mathematically, these different phases can be modeled by the sum of convex and concave functions \citep[e.g.][]{6788812}.  Then we can apply the classical Hermite-Hadamard (HH) inequality \citep[e.g.][]{rae/1149698532} to derive the relationship between $f(t_{m})$ and $f_{s}(t_{m})$. When the light curve is convex in a given time interval, the HH inequality states that $f(t_{m}) < f_{s}(t_{m})$. And we have $f(t_{m}) > f_{s}(t_{m})$ in the case of a concave light curve. In addition, if and only if the light curve is a linear function of time, the equality holds. In summary, we have 
\begin{itemize}
    \item $\Delta{f(t_{m})} > 0$ if $f(t_{m})$ is  convex,
    \item $\Delta{f(t_{m})} < 0$ if $f(t_{m})$ is  concave, and
    \item $\Delta{f(t_{m})} = 0$ if $f(t_{m})$ is linear. 
\end{itemize}
The detailed mathematical proof is provided in Appendix~\ref{app:hh}. It is observationally evident that the light curves of most real astronomical objects show non-linear dependence on time. Therefore, the time-averaged flux $f_{s}(t_{m})$ will be a biased measure of the true flux $f(t_{m})$. From the discussion above, it is expected that the $\Delta{t}$-bias tends to flatten the true light curve.  And the difference $\Delta{f(t_{m})}$ depends on the specific shape of the light curve and the exposure time as indicated by equation~(\ref{eq:diff}). 

Furthermore, we find that the $\Delta{t}$-bias can also alter the peak position. Assuming the peak positions are $t_\mathrm{peak}$ and $t_{s,\,\mathrm{peak}}$ for the true light curve and time-averaged light curve, respectively, the mathematical proof (see Appendix~\ref{app:ps}) indicates that their difference $\delta{t_\mathrm{peak}}$ (defined as $t_{s,\,\mathrm{peak}}-t_\mathrm{peak}$) always satisfies $|\delta{t_\mathrm{peak}}|<\Delta{t}/2$. For light curves with fast rise and then slow decay phases around the peak, $t_{s,\,\mathrm{peak}}$ will shift rightwards ($t_{s,\,\mathrm{peak}}>t_\mathrm{peak}$). While for light curves with slow rise and then fast decay phases around the peak, $t_{s,\,\mathrm{peak}}$ shifts leftwards ($t_{s,\,\mathrm{peak}}<t_\mathrm{peak}$). And if the light curve is axisymmetric with respect to $t_{m}=t_\mathrm{peak}$, the peak position does not change so that we have $t_{s,\,\mathrm{peak}}=t_\mathrm{peak}$. 

An illustration of the $\Delta{t}$-bias is shown in the top panel of Fig.~\ref{fig:tbias_demo}. The solid line represents an empirical light curve template of  white-light stellar flares constructed by using the 1-minute exposure data from \textit{Kepler} mission \citep{2022AJ....164...17M}. It is simple to prove that in the early rise and late decay phases, the light curve is convex, while it is concave around the peak. In this model, three free parameters are used to describe the general profile of a flare light curve: the peak time $t_\mathrm{peak}$,  flux amplitude $A$, and FWHM. For the default template as shown in the figure, we have $t_\mathrm{peak}=-0.0026$, $A=0.9502$, and FWHM=1.0867, respectively. Here we define the scale parameter as $\alpha=\Delta{t}/\mathrm{FWHM}$.  The time-averaged light curves derived by equation~(\ref{eq:dtFlux}) with different exposure times are  shown in the top panel of Fig.~\ref{fig:tbias_demo}. It can be seen that significant deviations present in the fast rise and peak phases between the true and time-averaged fluxes. In the late decay phase, the deviations become smaller because the light curves tend to be linear over time. The $\Delta{t}$-bias does make the profile of the light curve flatter as the increase of exposure time. In addition, as we have demonstrated,  as the increase of exposure time, not only does the amplitude of the peak get smaller, but the position of the peak systematically shifts rightwards. In the bottom panel of Fig.~\ref{fig:tbias_demo}, we show the differences of the  parameters $t_\mathrm{peak}$,  $A$, and FWHM relative to those of the template light curve as a function of exposure time. 

We note that the light curves of many other transient events, such as supernovae and fast blue optical transients \citep{2020ApJ...904...35P, 2023ApJ...949..120H}, also present the similar profile as the stellar flares, i.e. a fast rise phase and then a following slow decay phase. Therefore, the results described here are also expected to hold for them.

In addition to the bias induced by  exposure time, observing cadence can also affect the analysis of photometric light curves. The cadence does not alter the true flux, but imposes a finite rate of sampling, making the observed light curves discrete. Mathematically,  this discreteness can be described by a pointwise product of the true light curve $f(t)$  and a Dirac Comb window function $W_{\{t_{m,\,i}\}}(t)$, which is a sequence of Dirac delta functions spaced with a given time series \{$t_{m,\,i}$\} where $i$ ranges from 1 to $n$, and $n$ is the total number of exposures \citep{2018ApJS..236...16V,2020sdmm.book.....I}. Specifically, the discrete light curve can be written as 
\begin{equation}\label{eq:cadence}
\mathcal{F}(t) = f(t)\,W_{\{t_{m,\,i}\}}(t) = \sum_{i=1}^{n}f(t_{m,\,i})\,\delta(t-t_{m,\,i}),
\end{equation}
where $\delta(t)$ is the Dirac delta function at time $t$.
As we mentioned earlier, the cadence is usually nonuniform because of the influences of observing time windows and instrumental operations. In the context of spectral analysis, the cadence may result in aliasing effects that hinder the accurate quantification of the underlying true signals. For more discussions on this aspect, we direct interested readers to  \citet{2018ApJS..236...16V} and references therein. In Fig.~\ref{fig:tbias_demo}, the vertical dashed lines  illustrate a window function $W(t)$, and the circles depict the resulting discrete light curve for our flare template. In this example, the window function can be  expressed as $W_{\{t_{m,\,i}\}}(t) = \sum_{i=1}^{3}\delta(t-t_{m,\,i})$.

However, we emphasize that in the presence of $\Delta{t}$-bias, the cadence-induced pointwise product in equation~(\ref{eq:cadence}) should be modified. That is, the real observed light curve must contain effects due to both exposure time and cadence. Thus, we can substitute the true flux with the time-averaged flux that is obtained from equation~(\ref{eq:dtFlux}), and express the observed discrete light curve as
\begin{equation}\label{eq:expcad}
\mathcal{F}_{s}(t) = \sum_{i=1}^{n}f_{s}(t_{m,\,i})\,\delta(t-t_{m,\,i}).
\end{equation}
As illustrated in Fig.~\ref{fig:tbias_demo}, under the given window function $W_{\{t_{m,\,i}\}}(t)$ and exposure time $\Delta{t}$, the discrete light curve with $\alpha=1/2$ (i.e., $\Delta{t}=\mathrm{FWHM}/2$ as defined before)  is supposed to be observed as squares instead of circles.

Besides the impact on light curve morphology, $\Delta{t}$-bias can also distort color measurements. Astronomical color is defined as the magnitude difference between two filters. For filters $X$ and $Y$, the color $c_{XY}(t_{m})$ at time $t_{m}$ is calculated as $c_{XY}(t_{m})=m^{X}(t_{m}) - m^{Y}(t_{m})$, where $m^{X}(t_{m})$ and $m^{Y}(t_{m})$ are  magnitudes for the two filters. Conventionally, the wavelength of filter $X$ is shorter than that of $Y$. Because the true light curve of a transient or variable is usually color-dependent, and the exposure times for different filters vary in real observations, we can, without loss of generality, substitute equation~(\ref{eq:dtFlux}) into this color equation to yield 
\begin{align}\label{eq:dtColor}
c_{XY}(t_{m}) &= -2.5\log_{10}(f_{s}^{X}(t_{m})) + 2.5\log_{10}(f_{s}^{Y}(t_{m})) \nonumber \\
&= -2.5\log_{10}\left[\frac{\int_{t_{m}-\Delta{t^{X}}/2}^{t_{m}+\Delta{t^{X}}/2} f^{X}(t) \mathrm{d}t}{\int_{t_{m}-\Delta{t^{Y}}/2}^{t_{m}+\Delta{t^{Y}}/2} f^{Y}(t) \mathrm{d}t} \right] + 2.5\log_{10}(\frac{\Delta{t^{X}}}{\Delta{t^{Y}}}),
\end{align}
where $f^{X}(t)$ and $f^{Y}(t)$ denote the true light curves for filters $X$ and $Y$, respectively. The corresponding time-averaged fluxes are $f_{s}^{X}(t_{m})$ and $f_{s}^{Y}(t_{m})$, with exposure times of $\Delta{t^{X}}$ and $\Delta{t^{Y}}$. 
This equation reveals that the color also depends on the exposure time $\Delta{t}$. Even if $\Delta{t^{X}}=\Delta{t^{Y}}$, the first term remains a function of $\Delta{t}$. Furthermore, we note that observations in different filters generally have distinct cadences or window functions. To calculate the colors of a  transient or variable and deduce the associated physical properties, we must interpolate or model the light curves of different filters to a common time grid. Therefore, properly incorporating the $\Delta{t}$-bias will become necessary.

In summary, we stress that the $\Delta{t}$-bias induced by exposure time is  non-negligible for light curve analysis, particularly when the exposure time is comparable to the timescale of a target transient or variable.  Additionally, cadence can introduce extra computational complexity. To extract unbiased physical parameters, one feasible approach is to incorporate the $\Delta{t}$-bias and cadence directly into the theoretical light curves or colors, and then fit them to observational data.

\section{Case Studies}\label{sec:cases}
In this section, we further study the $\Delta{t}$-bias for the light curves of stellar flares, periodic stars, and AGNs.  In what follows, we assume that the exposure time $\Delta{t}$ is always constant and there are no additional overheads or other gaps between two sequential exposures.

\subsection{Stellar Flares}

\begin{figure}
    \centering
    \includegraphics[width=1.0\linewidth]{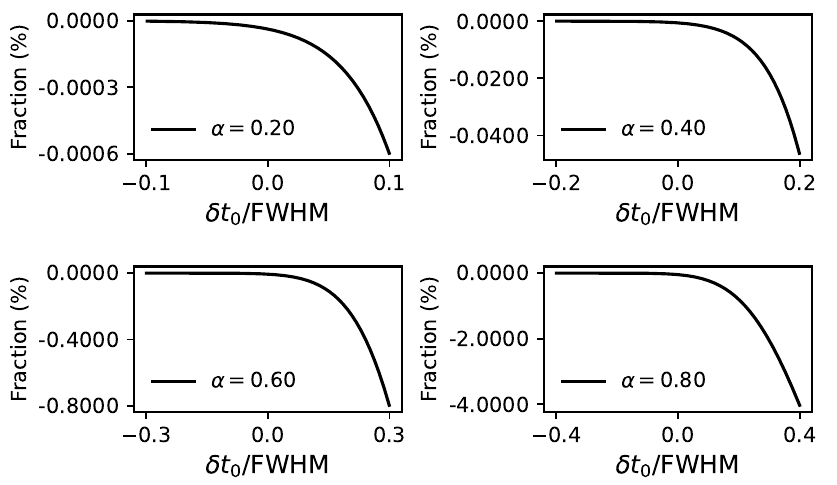}
    \caption{Fraction of the total flux as a function of relative time offset. The four panels show the results with different exposure times.}
    \label{fig:sf_energy}
\end{figure}

We start with the discussion of the empirical stellar flare template as shown in Fig.~\ref{fig:tbias_demo}. The initial eruption time of the stellar flare is at $t_{0}=-1$, and the total flux $f_\mathrm{tot}$ estimated by the integral under the light curve is 2.0409. Besides the aforementioned effects induced by the $\Delta{t}$-bias, we further investigate the total  integrated flux $f_{s,\,\mathrm{tot}}$ of the time-averaged light curve. 
Since the light curve is discrete, the trapezoidal rule is adopted to estimate the total flux, which is
\begin{equation}\label{eq:estimate_energy}
f_{s,\,\mathrm{tot}} = \sum_{i=1}^{n-1} \frac{f_s(t_{m,\,i}) + f_s(t_{m,\,i+1}) }{2}\Delta{t},   
\end{equation}
where $i$ is the $i$-th exposure and $n$ is the total number of exposures. Because exposures before $t_{0}$ do not contribute to the total flux estimate, we can unambiguously set  $t_{0}$ to be within the first exposure interval $[t_{m,\,1}-\Delta{t}/2,\,\, t_{m,\,1}+\Delta{t}/2]$, and define the relative offset as $\delta{t_{0}} = t_{m,\,1} - t_{0}$. Similarly, we denote the eruption end time as $t_{e}$, and set it to be within the last exposure interval $[t_{m,\,n}-\Delta{t}/2,\,\, t_{m,\,n}+\Delta{t}/2]$. The relative time offset between $t_{e}$ and $t_{m,\,n}$ is $\delta{t_{e}} = t_{m,\,n} - t_{e}$.
Immediately, we have the inequalities $|\delta{t_{0}}|< \Delta{t}/2$ and $|\delta{t_{e}}|< \Delta{t}/2$.  
Based on these definitions, we can  derive the relationship between $f_{s,\,\mathrm{tot}}$ and $f_\mathrm{tot}$ (see Appendix~\ref{app:fc}), which follows 
\begin{align}\label{eq:ftot_corr}
f_{s,\,\mathrm{tot}} &= f_\mathrm{tot} - \frac{1}{2}\left[f_{s}(t_{m,\,1}) + f_{s}(t_{m,\,n})\right]\Delta{t} \nonumber \\
&= f_\mathrm{tot} - \frac{1}{2}\left[\int_{t_{m,\,1}-\delta{t_{0}}}^{t_{m,\,1}+\Delta{t}/2}f(t)\mathrm{d}t + \int^{t_{m,\,n}-\delta{t_{e}}}_{t_{m,\,n}-\Delta{t}/2}f(t)\mathrm{d}t\right].
\end{align}
Equation~(\ref{eq:ftot_corr}) demonstrates that $f_{s,\,\mathrm{tot}} $ underestimates the true  total flux $f_\mathrm{tot}$. The degree of underestimation depends on the time offsets $\delta{t_{0}}$ and $\delta{t_{e}}$, the exposure time $\Delta{t}$, and the inherent profile of the light curve. Only under the extreme case that the initial eruption time of a stellar flare $t_{0}$ is completely coincident  with the end time of the first exposure (i.e., $t_{0} = t_{m,\,1} + \Delta{t}$/2) and the end time $t_{e}$ coincides with the start of the last exposure (i.e., $t_{e} = t_{m,\,n} - \Delta{t}$/2), $f_{s,\,\mathrm{tot}}$ can be an unbiased estimator of  $f_\mathrm{tot}$.  However, equation~(\ref{eq:ftot_corr}) also reveals that a simple correction factor $[f_{s}(t_{m,\,1}) + f_{s}(t_{m,\,n})]\Delta{t}/2$ can be added to $f_{s,\,\mathrm{tot}}$ to calculate the corrected total flux. Fig.~\ref{fig:sf_energy} illustrates the fraction of underestimation for different exposure times for the stellar flare template. In each of the four panels, the horizontal axis is the relative time offset $\delta{t_{0}}$ which is normalized to the FWHM, and the vertical axis is the flux fraction, defined as $f_{s,\,\mathrm{tot}}/f_\mathrm{tot} -1$. As a practical example,  \citet{2018ApJ...859...87Y} reported that the flare energies estimated from \textit{Kepler} long-exposure data are systematically underestimated by 25\% when compared to those obtained from short-exposure data. This can be partially explained by Fig.~\ref{fig:sf_energy}, which illustrates that the fraction of underestimation indeed increases with the increase of exposure time. In addition, equation~(\ref{eq:ftot_corr}) also indicates that accurately determining  the eruption start time $t_{0}$ and end time $t_{e}$ (or the flare duration $t_{e}-t_{0}$) is  crucial for unbiased total flux calculation. However, \citet{2018ApJ...859...87Y} found systematic differences in these parameters.  For more discussions on the underestimation, interested readers are referred to their work.

\begin{figure}
    \centering
    \includegraphics[width=1\linewidth]{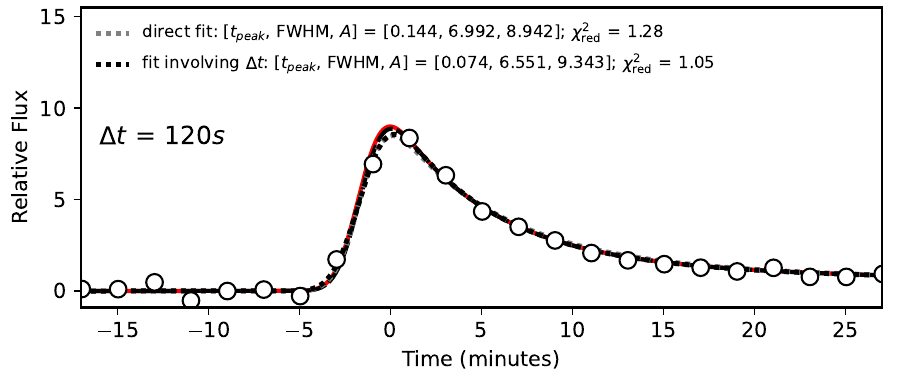}
    \includegraphics[width=1\linewidth]{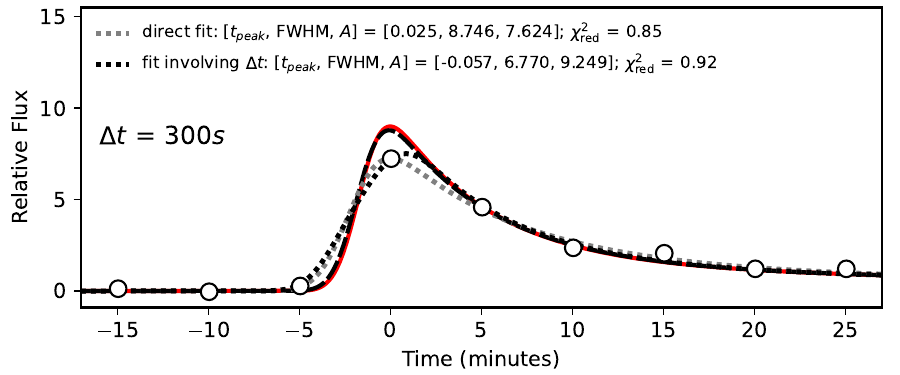}
    \includegraphics[width=1\linewidth]{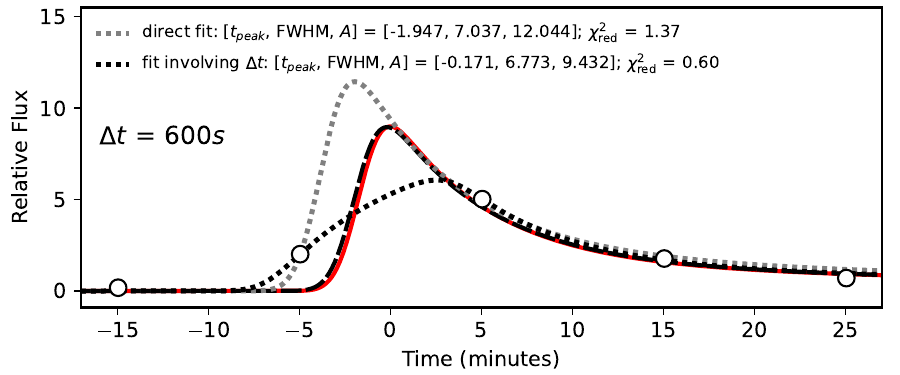}
    \caption{Illustration of the impact of $\Delta{t}$-bias on  stellar flare light curve fitting. In each panel, the solid red line is the input fiducial true light curve profile, and the circles represent the simulated light curve with exposure time $\Delta{t}$. The  gray dotted line is the result by fitting the simulated data using the parameterized stellar flare profile directly. The  black dashed line is the best-fit profile taking the $\Delta{t}$-bias into account, and the  black dotted line corresponds to the time-averaged profile (see text for details).}
    \label{fig:stellar_flare_fit}
\end{figure}

To further investigate the $\Delta{t}$-bias for practical light curve analysis,  we compare the light curve fitting results with and without taking the exposure time into account. We firstly simulate stellar  flare light curves with different exposure times. The parameters of the fiducial true  light curve profile are set to be  $A=9.469$ and $\mathrm{FWHM}=6.540\,\mathrm{minutes}$. These values represent the best-fit means  \citep{2024ApJS..271...57X} for the flare samples from \citet{2022ApJ...926..204H}. In addition, we arbitrarily set the peak time as $t_\mathrm{peak}=0.0\,\mathrm{minute}$. The red line in each panel of Fig.~\ref{fig:stellar_flare_fit} displays the fiducial true light curve profile. We then generate the time-averaged light curves using  equation~(\ref{eq:dtFlux}) with exposure times $\Delta{t}=120$\,s, $300$\,s, and $600$\,s, respectively. The noise of each light curve is assumed to follow the Gaussian distribution $N\sim(0, \,\sigma^2)$, where $\sigma=A/50$, meaning that the signal-to-noise ratio of the peak amplitude is fixed to 50. The final time-averaged light curves with different exposure times are shown as circles in Fig.~\ref{fig:stellar_flare_fit}. 

Next, we fit these simulated light curves. The gray dotted line in each panel of Fig.~\ref{fig:stellar_flare_fit} represents the result by fitting the data using the parameterized stellar flare profile directly. This is also the generally adopted method in current studies \citep[e.g.][]{2022AJ....164...17M,2024arXiv240407812V}. The best-fit parameters are provided in each panel and Table~\ref{tab:flare_fit}. Obviously, the best-fit parameters are not consistent with the fiducial values. On the other hand, in order to take the $\Delta{t}$-bias  into account, we substitute the parameterized stellar flare profile into equation~(\ref{eq:dtFlux}), i.e., $f(t)$ inside the integral, and derive the best-fit parameters again using the circled data points. The result is shown by the black dashed line in each panel of Fig.~\ref{fig:stellar_flare_fit}.  The black dotted line corresponds to the profile obtained by the time-averaging of the solid black line. It can be seen that the black dashed line matches the fiducial model  much better compared to the  gray dotted lines from the direct fitting, though the two methods can both fit the data well and give similar reduced $\chi^2_\mathrm{red}$. In addition, we note that the direct fitting method tends to give larger deviation from the fiducial model as the exposure time increases. The simulation results indicate that it is indispensable to consider the $\Delta{t}$-bias in the light curve fitting procedures in order to obtain unbiased estimate of the physical quantities.

\begin{figure}
    \centering
    \includegraphics[width=1\linewidth]{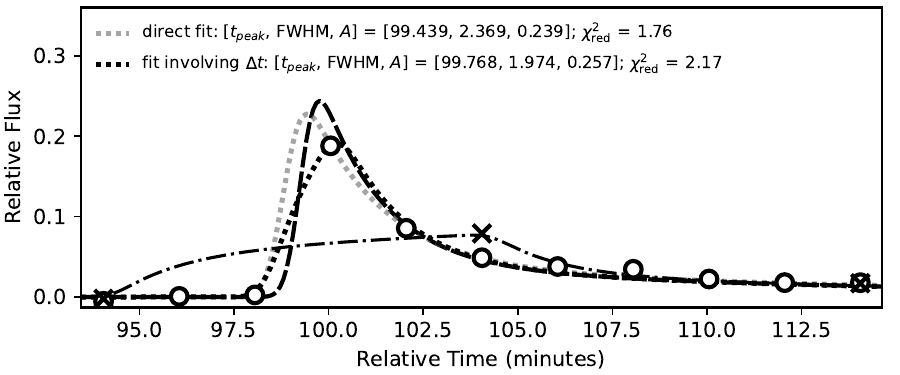}
    \caption{Light curve fitting of a  real stellar flare occurring on an M3.5 star (TIC 197829751) observed by  \textit{TESS}. The circles and crosses represent the  light curves with exposure times of $\Delta{t}$=120\,s, and 600\,s, respectively. Same as Fig.~\ref{fig:stellar_flare_fit}, the gray dotted line is the best-fit light curve  by directly fitting the 120s-exposure data. The black dashed and dotted lines are the best-fit profile when taking the $\Delta{t}$-bias into account, and the corresponding time-averaged profile, respectively. The dash-dotted line represents  the predicted 600s-exposure light curve.}
    \label{fig:stellar_flare_apply}
\end{figure}

We then apply the same fitting methods to a real stellar flare occurring on an M3.5 star (TIC 197829751) from   \textit{TESS} data \citep{2019AJ....157..234S}.  \textit{TESS} observed the stellar flare with three different exposure times: $\Delta{t}=20$\,s, $120$\,s, and $600$\,s. Since the stellar flare template is constructed using the 1-minute exposure  data from \textit{Kepler} mission, in principle it is not feasible to fit the light curve observed with shorter exposure due to the $\Delta{t}$-bias. This is evident from the poor fit to the peak fluxes as shown in  \citet[Fig. 10]{2022AJ....164...17M}. Thus we opt to fit only the 120s-exposure light curve, and show the best-fit results in Fig.~\ref{fig:stellar_flare_apply}. We note that the parameters estimated by the direct fitting method are identical to that provided by  \citet{2022AJ....164...17M}. Similar to the simulation, when involving the $\Delta{t}$-bias, the estimated parameters are  different, particularly for the FWHM which decreases from 2.369\,minutes to 1.974\,minutes. We then integrate the best-fit light curve profile (black dashed line) with the longest exposure time $\Delta{t}=600$\,s using  equation (\ref{eq:dtFlux}) to predict the corresponding light curve. It can be seen that the prediction (dash-dotted line) is completely consistent with the observed light curve with exposure time of $\Delta{t}=600$\,s.

\subsection{Periodic Stars}
For periodic stars, we assume that the true light curve varies periodically by following 
\begin{equation}\label{eq:tped}
f(t_{m}) = A\,\left[\sin(2\pi\omega t_{m} + \phi) + B\right],
\end{equation}
where $A$ and $B$ are constants, and required to satisfy $A>0$ and $B>1$ to ensure $f(t_{m})$ is always positive. In addition, $\omega$ is the frequency which is related to the period $T$ by  $T=1/\omega$, and $\phi$ is the phase. 
Based on the periodicity of the sine function, the light curve is concave in the interval $t_{m} \in [k/\omega-\phi/2\pi\omega, k/\omega-(\phi-\pi)/2\pi\omega]$ where $k$ is an integer starting from 0. And it is convex in the interval $[k/\omega-(\phi-\pi)/2\pi\omega, k/\omega-(\phi-2\pi)/2\pi\omega]$.
Inserting the above equation into equation~(\ref{eq:dtFlux}), we get
\begin{equation}\label{eq:taped}
    f_{s}(t_{m}) = A \, \left[\frac{\sin(\pi \omega \Delta{t})}{\pi \omega \Delta{t}}\sin(2\pi\omega t_{m} + \phi) + B\right].
\end{equation}

\begin{figure}
    \centering
    \includegraphics[width=1\linewidth]{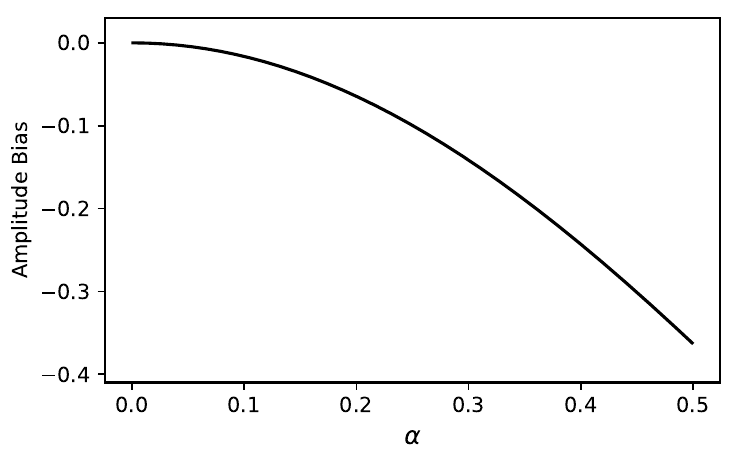}
    \caption{Amplitude bias as a function of relative exposure time for periodic stars with sinusoidal light curve.}
    \label{fig:period}
\end{figure}

By comparing equations~(\ref{eq:tped}) and ~(\ref{eq:taped}), it is found that the period and position of the peak are not affected by the $\Delta{t}$-bias. In addition, the average flux can be calculated by integrating the light curve over a period $T$. By definition, the average flux of the true light curve $f(t_{m})$ is $\int_{0}^{T}f(t_{m})\mathrm{d}t_{m}/T = AB/T$, and the same value holds for the time-averaged light curve $f_{s}(t_{m})$. That is, the average flux (or equivalently, the average magnitude) is also unchanged. But an extra term ${\sin(\pi \omega \Delta{t})}/{(\pi \omega \Delta{t})}$ appears in the time-averaged flux $f_{s}(t_{m})$, meaning that the amplitude gets smaller since ${\sin(\pi \omega \Delta{t})}/{(\pi \omega \Delta{t})}<1$ always holds when $\Delta{t}>0$. By defining the relative time scale $\alpha=\Delta{t}/T=\omega\Delta{t}$,  the dependence of the amplitude bias on the parameter $\alpha$, which is $\sin(\pi \alpha)/(\pi \alpha) - 1$, is presented in Fig.~\ref{fig:period}. As seen in the figure, it is expected that the $\Delta{t}$-bias will be significant for the study of short period stars with periods of several minutes or shorter \citep{2023A&A...669A..48B,2024arXiv240302384B}. 

One of the important applications of periodic stars is to determine cosmic distances, such as through the period-color-luminosity relation based on the Cepheid or RR Lyrae variables \citep[e.g.][]{2020JApA...41...23B}. Generally, the color and luminosity of such a variable are derived from the average magnitudes of the light curves obtained using different filters, through Fourier decomposition or template fitting \citep{1971A&AS....4..265S,2010ApJ...708..717S}. As we have proved, both the period and the average magnitude remain unaffected by the $\Delta{t}$-bias. In principle, it is expected that the period-color-luminosity relation is also unaffected. However, it becomes necessary to account for the $\Delta{t}$-bias when we investigate any relation related to amplitude, such as the period-amplitude relation \citep[e.g.][]{2020JApA...41...23B}. 

In addition, it is worth further discussing the template fitting method. It has been widely applied in deriving the average magnitude and amplitude of a light curve \citep{2010ApJ...708..717S, 2022ApJ...938..101S}. In particular, when the light curve is sparsely sampled, the template fitting method is more reliable \citep{2007A&A...462.1007K}.
However, these template light curves are constructed from observations with specified exposure times and cadences, which already incorporate the $\Delta{t}$-bias. When fitting these templates to light curves, which are usually collected from different observations with distinct exposure times and cadences, potential biases may be introduced into the derived parameters. Detailed investigations are not within the scope of this paper, and we leave them for future studies.

\subsection{AGNs}
The light curves of  AGNs can be described by the damped random walk model \citep{2009ApJ...698..895K}. For short timescales, the power spectral density (PSD) of the AGN light curves follows $\mathrm{PSD}(f) \propto f^{-\alpha}$, where $f$ is the frequency, and $\alpha$ is the power-law index. In the case of a random walk process, we have $\alpha=2$. Another way to model the variability of AGNs is the structure function (SF) which is defined as the flux (or magnitude) difference as a function of time lag $\delta{t}$ \citep[e.g.][]{2016ApJ...826..118K}:
\begin{equation}
\mathrm{SF}(\delta{t}) = \sqrt{\frac{1}{N_{\delta{t}, \mathrm{pair}}} \sum_{i=1}^{N_{\delta{t}, \mathrm{pair}}} \left(f(t) - f(t+\delta{t})\right)^2 },
\end{equation}
where $N_{\delta{t}, \mathrm{pair}}$ is the number of pairs for a given $\delta{t}$, and $f$ is the flux. For short timescale, the SF also follows a power-law form $\mathrm{SF(\delta{t})} \propto \delta{t}^{\gamma}$, and $\gamma=0.5$ for the damped random walk model. Here we investigate how the $\Delta{t}$-bias affects the determination of the power-law index $\gamma$ of the SF. 

\begin{figure}
    \centering
    \includegraphics[width=1\linewidth]{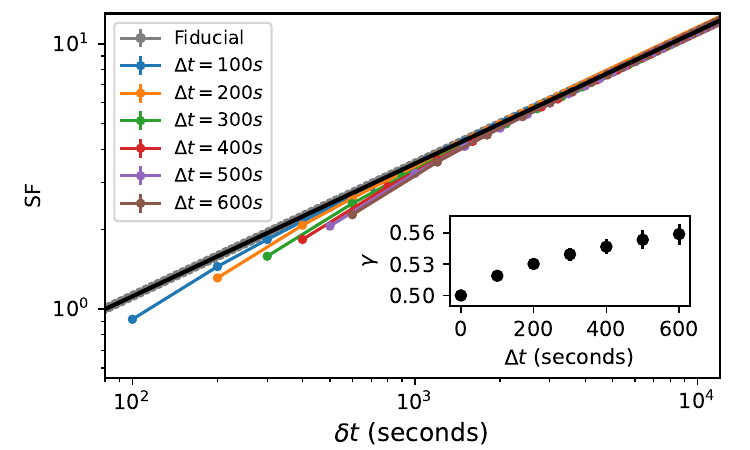}
    \caption{Comparison of the SFs with different exposure times. For a given exposure time, the SF is an average of 100 individual SFs. The gray dots with error bars and black line represent the fiducial and expected true SF with $\gamma=0.5$, respectively. The power-law index $\gamma$ as a function of exposure time is shown in the inset.}
    \label{fig:agn_sf}
\end{figure}

We use \texttt{Stingray}, a toolkit developed by \citet{2019ApJ...881...39H}, to generate the simulated AGN light curves by following the random walk process. The mean and standard deviation of the fiducial light curves are set to be 100.0 and 0.5, respectively. In addition, since the time resolution cannot be infinitesimal, we set it to be 0.1 seconds. In total, 100 fiducial light curves are generated with different random seeds. We then get the light curves with different exposure times by following equation~(\ref{eq:dtFlux}) and further calculate the corresponding SFs. For a given exposure time $\Delta{t}$, we average the 100 individual SFs and show the result in Fig.~\ref{fig:agn_sf}. It can be seen that the SFs with different exposure times are all steeper compared to the fiducial SF (gray dots with error bars) and the expected true SF with $\gamma=0.5$ (black line) at short timescale. The steepness becomes larger as the increase of exposure time. We can also find that the SFs are consistent with the fiducial one at long timescale, indicating that the $\Delta{t}$-bias mostly affects the short-timescale variability of AGNs. To quantify the impact of $\Delta{t}$-bias on the SFs, we fit the averaged SFs with the power-law formula. The upper limit of the fitting is fixed to 0.1\,days. The best-fit power-law index $\gamma$ as a function of exposure time is shown in the inset of Fig.~\ref{fig:agn_sf}. Remarkably, the $\gamma$ value increases about 12\% when the exposure time is 600\,s. The results indicates that the explanation about  the  variability features of AGNs at short timescale should be careful in the presence of $\Delta{t}$-bias, particularly for the study of intra-night variability of AGNs \citep[][and references therein]{1995ARA&A..33..163W}. 

\section{Discussions and Conclusions}\label{sec:conc}
In this paper, we have analyzed the impacts of the bias induced by exposure time on observed photometric light curves. Mathematical calculations showed that the time-averaged flux with a given exposure time is not  necessarily identical to the expected value of the true light curve. The bias or their difference depends on the duration of the exposure time and the specific shape of the true light curve. Generally, when the true light curve is convex in a certain exposure, the time-averaged flux will be larger than the expected true flux. It is smaller when the  true light curve is concave. As a result, the exposure time tends to flatten the true light curve and enlarge the characteristic timescale (e.g. the FWHM). Meanwhile, the peak position and photometric color are also altered.

We have discussed the bias for three types of light curves in detail: stellar flares, periodic stars, and AGNs. For the stellar flares, like many other transients, their light curves present three remarkable phases, a fast rise phase, a sharp peak phase, and then a  slow decay phase. Besides the systematic underestimation of the peak flux and shift of the peak position, the exposure time bias can also cause the total integrated flare flux to be underestimated. The fraction depends on the duration of the exposure time. We further demonstrated the impact of the bias on the morphological parameters that were estimated by fitting both simulated and real observed light curves. For periodic stars, our analysis indicates that the bias will not affect the period and peak position determination, but can give rise to an underestimate of  the peak flux. For the AGNs, we calculated the SFs of a set of simulated light curves that follow the random walk process and demonstrated that the exposure time can make the slopes of the SFs steeper at short timescale. It turns out that  the steepness is proportional to the exposure time. In addition to the analysis presented here, however, it will be interesting to further explore the impacts of the exposure time bias  on other aspects regrading light curve studies, and extend the analysis to many other transients and variables.  Moreover, it usually takes even longer exposure times to perform spectroscopic observations for these transients and variables. Thus it is expected that quantifying the bias for the derived parameters from the spectra is also necessary. 

We emphasize that the bias is sensitive to the exposure time or the relative exposure time, i.e. the ratio of the exposure time to the typical timescale of the light curve. For variables or transients with long timescale variation, such as  stars with period longer than days or supernovae with typical timescale of weeks, the relative exposure time is very small (typically $\Delta{t}/\omega_{t}\lesssim 10^{-3}$), and thus the bias is expected to be negligible. However, with the increase of the discoveries of the short period stars and fast transients \citep{2021OJAp....4E..14D,2023PASP..135j5002H,2023A&A...669A..48B,2024arXiv240302384B},  it is essential to account for the bias if  unbiased physical parameters are desired. 

\section*{Acknowledgements}
We thank the anonymous referee for the constructive comments that significantly enhanced the quality of our manuscript. Dezi Liu acknowledges the supports from NSFC under the grant of 12103043, and National Key R\&D Program of China No. 2022YFF0503404. Zuhui Fan is
supported by NSFC under 11933002, U1931210 and 11333001.

This paper includes data collected with the TESS mission, obtained from the MAST data archive at the Space Telescope Science Institute (STScI). Funding for the TESS mission is provided by the NASA Explorer Program. STScI is operated by the Association of Universities for Research in Astronomy, Inc., under NASA contract NAS 5–26555.  
This research made use of Lightkurve, a Python package for Kepler and TESS data analysis \citep{2018ascl.soft12013L}. 
This work made use of Astropy:\footnote{http://www.astropy.org} a community-developed core Python package and an ecosystem of tools and resources for astronomy \citep{2013A&A...558A..33A,2018AJ....156..123A,2022ApJ...935..167A}.
Besides, this work has also used the following Python packages: \texttt{numpy} \citep{harris2020array}, \texttt{scipy} \citep{2020SciPy-NMeth}, \texttt{matplotlib} \citep{Hunter:2007}, and \texttt{Stingary} \citep{2019ApJ...881...39H}.

\section*{Data Availability}
The data underlying this article will be shared on reasonable request to the corresponding authors.


\bibliographystyle{mnras}
\bibliography{timeBias} 


\appendix
\section{Hermite-Hadamard inequality}\label{app:hh}
The Hermite-Hadamard inequality states that for a given convex function $g(x)$, in the interval $[a, b] \rightarrow \mathbb{R}$, we have 
\begin{equation}
    g(\frac{a+b}{2}) \leq \frac{1}{b-a}\int_{a}^{b}g(x)\mathrm{d}x.
\end{equation}
More complete form and relevant proof can be found in \citet{rae/1149698532} and references therein. Similarly, if $g(x)$ is a concave function, then 
\begin{equation}
    g(\frac{a+b}{2}) \geq \frac{1}{b-a}\int_{a}^{b}g(x)\mathrm{d}x.
\end{equation}
In both cases, equality holds only for functions of the form $g(x) = kx+d$, where $k$ and $d$ are real numbers.

Substituting equation~(\ref{eq:cflux}) and equation~(\ref{eq:dtFlux}) into the Hermite-Hadamard inequality, it is straightforward to prove that when the light curve is convex, we have 
\begin{align}
    f(t_{m}) &= f(\frac{t_{m}-\Delta{t}/2 + t_{m}+\Delta{t}/2}{2}) \nonumber \\
    &\leq \frac{1}{\Delta{t}} \int_{t_{m}-\Delta{t}/2}^{t_{m}+\Delta{t}/2}f(t)\mathrm{d}t \nonumber \\
    &= Z_{f} \frac{1}{\Delta{t}} \int_{t_{m}-\Delta{t}/2}^{t_{m}+\Delta{t}/2}\mathrm{d}t\int_{0}^{\infty}s_{\nu}(t)R_{\nu}\frac{\mathrm{d}\nu}{\nu} \nonumber \\
    &= f_{s}(t_{m}), 
\end{align}
i.e., $f(t_{m})\leq f_{s}(t_{m})$.  When the light curve is concave, it can be demonstrated that  $f(t_{m})\geq f_{s}(t_{m})$. Again, the equality holds only in the case of $f(t_{m})=kt_{m} +d$, namely, when the light curve is a linear function of time.

\section{Peak position shift}\label{app:ps}
For a  light curve with a single peak, we assume that the peak positions of  $f(t_{m})$ and $f_{s}(t_{m})$ are $t_\mathrm{peak}$ and $t_{s,\,\mathrm{peak}}$, respectively.  To quantify the peak position shift of $t_{s,\,\mathrm{peak}}$ with respect to $t_\mathrm{peak}$, we only need to compare their difference. In the following analysis, we define the difference as $\delta = t_\mathrm{peak} - t_{s,\,\mathrm{peak}}$. 

Because  the derivatives of $f(t_{m})$ and $f_{s}(t_{m})$ with respect to $t_{m}$ are both equal to zero at the peak positions. Namely, we have  equations 
\begin{equation}\label{eq:d2}
\left.\frac{\mathrm{d}f(t_{m})}{\mathrm{d}t_{m}}\right\vert_{t_{m}=t_\mathrm{peak}} = 0,
\end{equation}
\begin{equation}\label{eq:d1}
\left.\frac{\mathrm{d}f_{s}(t_{m})}{\mathrm{d}t_{m}}\right\vert_{t_{m}=t_{s,\,\mathrm{peak}}} = 0.
\end{equation}

We first discuss the case as shown in  Fig.~\ref{fig:tbias_demo}. The true light curve profile $f(t_{m})$ around the  peak $t_\mathrm{peak}$ is concave, and the absolute value of the derivative on the left of the peak is larger than that on the right, namely,
\begin{equation}
\left.\frac{\mathrm{d}f(t_{m})}{\mathrm{d}t_{m}}\right\vert_{t_{m}<t_\mathrm{peak}} > - \left.\frac{\mathrm{d}f(t_{m})}{\mathrm{d}t_{m}}\right\vert_{t_{m}>t_\mathrm{peak}}.
\end{equation}
If $\varepsilon$ is a positive value, then the above inequality yields  
\begin{equation}\label{eq:d4}
f(t_\mathrm{peak} - \varepsilon)  <  f(t_\mathrm{peak}+\varepsilon).
\end{equation}

In addition, substituting  equation~(\ref{eq:dtFlux}) into equation~(\ref{eq:d1}), we can derive 
\begin{equation}\label{eq:d3}
f(t_{s,\,\mathrm{peak}}-\Delta{t}/2) = f(t_{s,\,\mathrm{peak}}+\Delta{t}/2).
\end{equation}
Obviously, the equation only holds in the case of $t_{s,\,\mathrm{peak}}-\Delta{t}/2 < t_\mathrm{peak}$ and $t_{s,\,\mathrm{peak}}+\Delta{t}/2>t_\mathrm{peak}$. It means that $\delta$ should satisfy 
\begin{equation}
-\Delta{t}/2 < \delta < \Delta{t}/2.
\end{equation}

Combing the equations~(\ref{eq:d4}) and (\ref{eq:d3}), we further have 
\begin{equation}
f(t_\mathrm{peak}-\delta + \Delta{t}/2) = f(t_\mathrm{peak}-\delta - \Delta{t}/2) < f(t_\mathrm{peak}+\delta + \Delta{t}/2) \nonumber
\end{equation}
Because both $t_\mathrm{peak}-\delta + \Delta{t}/2$ and $t_\mathrm{peak}+\delta + \Delta{t}/2$ are larger than $t_\mathrm{peak}$, immediately we get $t_\mathrm{peak}-\delta + \Delta{t}/2 > t_\mathrm{peak}+\delta + \Delta{t}/2$, i.e. $\delta<0$. In summary, we have demonstrated  that $\delta$ satisfies 
\begin{equation}
-\Delta{t}/2 < \delta < 0.
\end{equation}
The above inequality indicates that the peak position of $f_{s}(t_{m})$ shifts rightwards with respect to the true peak position. And the offset is smaller than half of the exposure time.

Similarly, if the true light curve profile $f(t_{m})$ around the peak $t_\mathrm{peak}$ is axisymmetric, e.g., the sine function, it is easy to prove that $\delta = t_\mathrm{peak} - t_{s,\,\mathrm{peak}} = 0$. If the rise phase before the peak is relatively flatter than the decay phase as opposed to the light curve shown in Fig.~\ref{fig:tbias_demo}, the absolute value of the derivative on the left of the peak is expected to be smaller than that on the right. Following the same procedure, we can prove that 
 $0< \delta < \Delta{t}/2$. It means that the peak position of $f_{s}(t_{m})$ shifts leftwards with respect to the true peak position. And the offset is also smaller than half of the exposure time.

\section{Total Flux Estimation of stellar flare}\label{app:fc}
Theoretically, the total integrated flux of a stellar flare is calculated through
\begin{equation}
f_\mathrm{tot} = \int_{t_{0}}^{t_{e}}f(t)\mathrm{d}t, 
\end{equation}
where $t_{0}$ and $t_{e}$ denote the eruption start time and end time, respectively.
Observationally, we can approximate $f_\mathrm{tot}$ by equation~(\ref{eq:estimate_energy}) where the time-averaged flux at $t_{m,\,i}$ is 
\begin{equation}
f_{s}(t_{m,\,i}) = \frac{1}{\Delta{t}}\int_{t_{m,\,i}-\Delta{t}/2}^{t_{m,\,i}+\Delta{t}/2} f(t) \mathrm{d}t.
\end{equation}
To quantify the difference between  $f_\mathrm{tot}$ and $f_{s,\,\mathrm{tot}}$, we can expand  equation~(\ref{eq:estimate_energy}) as
\begin{align}
f_{s,\,\mathrm{tot}} &= \sum_{i=1}^{n-1} \frac{f_s(t_{m,\,i}) + f_s(t_{m,\,i+1}) }{2}\Delta{t}  \nonumber \\
&= \Delta{t}\left[\frac{1}{2} f_{s}(t_{m,\,1}) + f_{s}(t_{m,\,2}) + ... + f_{s}(t_{m,\,n-1}) + \frac{1}{2} f_{s}(t_{m,\,n})\right] \nonumber \\
&= \Delta{t}\sum_{i=1}^{n}f_{s}(t_{m,\,i}) - \frac{1}{2}\left[f_{s}(t_{m,\,1}) + f_{s}(t_{m,\,n})\right]\Delta{t}.
\end{align}
Under the assumption that there are no overheads or other gaps between two sequential exposures, we have $t_{m,\,i}+\Delta{t}/2 = t_{m,\,i+1}-\Delta{t}/2$. Therefore, combining the above three equations we can yield
\begin{equation}
f_{s,\,\mathrm{tot}} = f_\mathrm{tot} - \frac{1}{2}\left[f_{s}(t_{m,\,1}) + f_{s}(t_{m,\,n})\right]\Delta{t}.
\end{equation}

As we have defined, the first exposure interval is $[t_{m,\,1}-\Delta{t}/2,\,\, t_{m,\,1}+\Delta{t}/2]$ and the relative offset between $t_{0}$ and $t_{m,\,1}$ is $\delta{t_{0}} = t_{m,\,1} - t_{0}$. We expect that only a flare between $t_{0}$ and $t_{m,\,1}+\Delta{t}/2$ accounts for the flux budget of the first exposure.  Likewise, for the last exposure, only flare between $t_{m,\,n}-\Delta{t}/2$ and $t_{e}$ matters. Therefore, we finally have 
\begin{equation}
f_{s,\,\mathrm{tot}} = f_\mathrm{tot} - \frac{1}{2}\left[\int_{t_{m,\,1}-\delta{t_{0}}}^{t_{m,\,1}+\Delta{t}/2}f(t)\mathrm{d}t + \int^{t_{m,\,n}-\delta{t_{e}}}_{t_{m,\,n}-\Delta{t}/2}f(t)\mathrm{d}t\right].
\end{equation}
The equation indicates that the presence of the relative offsets between $t_{0}$ and $t_{m,1}$ and between $t_{e}$ and $t_{m,n}$ can cause potential underestimation of the total flux. Only under the extreme case that $t_{0}=t_{m,\,1}+\Delta{t}/2$ and $t_{e}=t_{m,\,n}-\Delta{t}/2$, we have $f_{s,\,\mathrm{tot}} = f_\mathrm{tot}$.
 
 \section{Stellar flare fitting results}\label{app:fs_fit}
The best-fit parameters of the simulated stellar light curves are shown in Table~\ref{tab:flare_fit}.
\begin{table}
\caption{Best-fit parameters of the simulated stellar flare light curves.}
\label{tab:flare_fit}
\begin{tabular}{clccc}
\hline
$\Delta{t}$              &  method       & $t_\mathrm{peak}$     & $\mathrm{FWHM}$  &      $A$ \\
(s)                           &                     &          (minutes)           &     (minutes)                      &        \\
\hline
--                              &     fiducial model                 &   0.000                 &    6.540                  &     9.469      \\
\hline
120                  & direct fit                       &   0.144                 &    6.992               &    8.942         \\
                        & $\Delta{t}$ involved     &  0.074                 &   6.551                 &    9.343                  \\
\hline
300                  & direct fit                      &   0.025                 &    8.746                  &    7.624         \\
                        & $\Delta{t}$ involved   &   -0.057  &   6.770                 &    9.249                  \\
\hline
600                  & direct fit                      &   -1.947                 &    7.037                  &    12.044         \\
                        & $\Delta{t}$ involved   & -0.171  &   6.773                 &    9.432                 \\
\hline
\end{tabular}
\end{table}

\bsp	
\label{lastpage}
\end{document}